\documentclass[conference]{IEEEtran}
\IEEEoverridecommandlockouts
\usepackage{cite}
\usepackage{amsmath,amssymb,amsfonts}
\usepackage{algorithmic}
\usepackage{booktabs}
\usepackage{graphicx}
\usepackage{pgfplots}
\pgfplotsset{compat=1.18}
\usepackage{textcomp}
\usepackage{xcolor}
\usepackage{comment}
\usepackage{tikz}
\usetikzlibrary{calc}
\usepackage{textgreek}
\usepackage[hidelinks]{hyperref}
\usepackage{listings}
\usepackage{subcaption}

\usepackage{titlesec}
\titlespacing*{\section}{0pt}{0.5\baselineskip}{0.2\baselineskip}
\titlespacing*{\subsection}{0pt}{0.3\baselineskip}{0.1\baselineskip}

\def\BibTeX{{\rm B\kern-.05em{\sc i\kern-.025em b}\kern-.08em
    T\kern-.1667em\lower.7ex\hbox{E}\kern-.125emX}}

\definecolor{loggray}{HTML}{F5F5F5}
\lstdefinestyle{terminal}{
    basicstyle=\ttfamily\scriptsize,
    backgroundcolor=\color{loggray},
    frame=single,
    rulecolor=\color{black!30},
    breaklines=true,
    columns=fullflexible,
    keepspaces=true,
    aboveskip=4pt,
    belowskip=4pt,
}

\definecolor{loggreen}{HTML}{2E7D32}
\definecolor{logred}{HTML}{C62828}
\definecolor{logorange}{HTML}{E65100}
\lstdefinestyle{serverlog}{
    basicstyle=\ttfamily\scriptsize,
    backgroundcolor=\color{loggray},
    frame=single,
    rulecolor=\color{black!30},
    breaklines=true,
    columns=fullflexible,
    keepspaces=true,
    aboveskip=4pt,
    belowskip=4pt,
}

\begin{document}

\title{Chained Attacks on Drone-Based Federated Learning: 
From Network Disruption to Device Impersonation}

\author{
\IEEEauthorblockN{Suleiman Muhammad Sabo, Hamed Alkharsh, Peilin Li, Chuadhry Mujeeb Ahmed,\\
Aydin Abadi, Shishir Nagaraja, Rajiv Ranjan}
\IEEEauthorblockA{School of Computing, Newcastle University, Newcastle upon Tyne, United Kingdom}
}

\maketitle

\begin{abstract}
Edge Intelligence (EI) has emerged as a transformative model for mission-critical unmanned platforms, such as drone swarms, by enabling collaborative model training at the network periphery. However, the security of FL deployments depends on both network availability and robust client authentication mechanisms. This paper investigates a chained attack against drone-based FL systems that combines network-layer denial-of-service with credential-based impersonation. We demonstrate that an adversary can: (1) force legitimate drones offline using 802.11 deauthentication attacks, and (2) subsequently impersonate the disconnected drone using extracted credentials. Through a systematic literature review and empirical validation using the Flower framework on two distinct testbeds of Raspberry Pi and Jetsons, we quantify the impact of availability disruptions under Independent and Identically Distributed (IID) and Non-Independently and Identically Distributed (Non-IID) data distributions, and confirm that single-factor authentication permits post-disconnect impersonation. Our findings reveal that even short-term wireless interruptions cascade into substantial training instability, particularly under non-IID conditions, while the authentication gap enables adversaries to seamlessly replace disconnected nodes. We discuss the compounded implications for mission-critical drone deployments and outline directions for future defenses addressing both availability and authentication vulnerabilities.
\end{abstract}


\section{Introduction}

Edge Intelligence (EI) and Federated Learning (FL) enable distributed devices to collaboratively train machine learning models without sharing sensitive raw data~\cite{fei_unmanned_2025,mcmahan2017fedavg, FLTA2025_NCL}. This decentralized paradigm is increasingly critical for Unmanned Aerial Vehicle (UAV) swarms operating in mission-critical environments, allowing for near-source data processing and real-time model adaptation~\cite{nguyen2021federated}. However, the distributed architecture of FL introduces severe security challenges regarding network availability and client authentication. 

In wireless edge environments, nodes are susceptible to 802.11 deauthentication (DeAuth) attacks that exploit unencrypted management frames to forcibly disconnect legitimate nodes~\cite{hamroun2025intrusion}. Concurrently, the integrity of FL relies entirely on authorized client participation~\cite{mothukuri2021survey}; if an adversary successfully impersonates a trusted client, they can inject malicious updates or compromise the global model. 

This paper examines how availability disruption and identity spoofing can be combined into a practical chained attack against drone-based FL systems. We demonstrate that an adversary can execute a two-stage attack: first, forcing a legitimate drone offline via network-layer deauthentication to induce immediate training degradation; second, exploiting the resulting session vacancy to seamlessly impersonate the disconnected drone using extracted credentials.

To empirically validate this threat, we move beyond software simulations and deploy a dual-platform physical testbed composed of Raspberry Pi and NVIDIA Jetson edge devices, utilizing the Flower framework~\cite{beutel2020flower}. 

Our key contributions are: \textbf{1)} The design and implementation of a compute-surrogate testbed to evaluate FL robustness under real-world network-layer deauthentication and impersonation attacks. \textbf{2)} Empirical quantification of FL accuracy degradation under DeAuth attack intensities up to 80\%, comparing IID and non-IID data distributions. \textbf{3)} Identification of a targeted ``expert node'' vulnerability under non-IID conditions, demonstrating that disconnecting data-concentrated clients causes disproportionate, class-specific model degradation. \textbf{4)} Confirmation that standard single-factor node authentication permits undetected, post-disconnect impersonation.

\section{Background and Related Work}
\label{sec:related-work}

\subsection{Authentication Vulnerabilities in Drone-Based FL}

UAVs and drone swarms operate in dynamic environments, relying on lightweight authentication schemes such as HMAC, blockchain, or cryptographic key exchanges~\cite{ouadah2024securing,salam2023seltha,mishra2023blockchain,akram2022bciod,aydin2022authentication,han2024scalable}. While specialized hardware~\cite{roy2024fpga} can increase assurance, it is rarely deployed in commercial swarms due to constraints on battery and computation~\cite{kammoun2024novel}. Consequently, a critical vulnerability across both general UAV networks~\cite{seo2023efficient,mahmood2025privacy,wang2024survey} and Edge-based Federated Learning (FL)~\cite{kairouz2021advances,ji2023lafed,huang2024eppafl,li2025enhancing,badhib2025iot} is the reliance on single-factor authentication. Most protocols treat the possession of valid credentials (e.g., cryptographic keys) as sufficient proof of identity~\cite{kumar2022secure,ju2024blockchain}. Because these frameworks lack Multi-Factor Authentication (MFA), they remain uniquely vulnerable to physical capture; an adversary can extract stored secrets from a downed drone and seamlessly impersonate it, bypassing network-layer protections entirely.

\subsection{Availability Attacks and the Research Gap}

While the FL literature extensively addresses benign ``stragglers'' and proposes dynamic subnetwork assignments for resource-constrained devices~\cite{feng2025practical,wen2022federated,listars,malandrino2021federated}, there is limited empirical research on targeted availability shocks. Recent surveys~\cite{cooray2025deep,janardhanan2025federated,almeida2025federated} identify communication challenges but do not empirically quantify the connectivity-accuracy tradeoffs under adversarial conditions.

Previous research largely treats network-level attacks and admission control in isolation. This work bridges that gap by demonstrating a chained attack that compounds 802.11 deauthentication with post-capture credential spoofing. While these individual primitives are established, our contribution lies in quantifying their combined impact on FL training dynamics---particularly the previously unexamined interaction with non-IID data heterogeneity. Table~\ref{tab:related_work} positions our contribution relative to the state-of-the-art.

\begin{table}[t]
\centering
\caption{Comparison of Related Work}
\label{tab:related_work}
\resizebox{\columnwidth}{!}{%
\begin{tabular}{@{}lccccc@{}}
\toprule
\textbf{Work} & \textbf{Domain} & \begin{tabular}{@{}c@{}}\textbf{Physical}\\\textbf{Capture}\end{tabular} & \begin{tabular}{@{}c@{}}\textbf{Availability}\\\textbf{Attack}\end{tabular} & \begin{tabular}{@{}c@{}}\textbf{Auth}\\\textbf{Bypass}\end{tabular} & \textbf{Empirical} \\
\midrule
Ouadah \& Merazka~\cite{ouadah2024securing} & UAV & No & No & No & No \\
Han et al.~\cite{han2024scalable} & Swarm & Partial & No & No & No \\
Mahmood et al.~\cite{mahmood2025privacy} & UAV/6G & Yes & No & Discussed & No \\
Ji et al.~\cite{ji2023lafed} & FL & No & No & No & Yes \\
Wen et al.~\cite{wen2022federated} & FL & No & Straggler only & No & Yes \\
\textbf{This work} & Drone FL & \textbf{Yes} & \textbf{Yes (DeAuth)} & \textbf{Yes} & \textbf{Yes} \\
\bottomrule
\end{tabular}%
}
\end{table}

\section{Threat Model and Attack Chain}
\label{sec:threat-model-and-attack-chain}

\subsection{Adversary Capabilities}

In our model, the attacker does not possess prior knowledge of the network architecture but instead employs active and passive reconnaissance to map the swarm's topology. This adversary model is characterized by two primary capabilities, which were validated on our testbed.

\begin{figure}[b]
    \centering
    \includegraphics[width=\columnwidth]{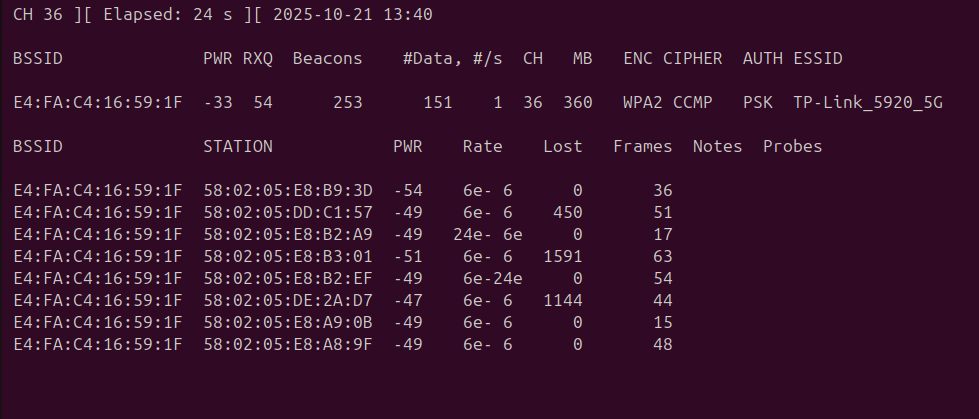}
    \caption{Targeted reconnaissance of the swarm BSSID, mapping specific station MAC addresses.}
    \label{fig:targetedscan}
\end{figure}

\textbf{Network Mapping and Role Identification.} The adversary utilizes a wireless interface in monitor mode to identify the Service Set Identifier (SSID) and Basic Service Set Identifier (BSSID) of the swarm. As shown in Fig.~\ref{fig:targetedscan}, a targeted scan of the identified network reveals the MAC addresses of all active stations.

\textbf{Credential Acquisition.} The adversary holds valid cryptographic credentials for at least one target node. We do not limit our threat model to a single acquisition method; instead, we recognize credential compromise as a realistic precondition that can occur through various means, including physical capture and forensic extraction from downed drones, supply chain infiltration, insider access, or insecure over-the-air provisioning. This assumption is consistent with recent analyses of unmanned edge intelligence, which identify credential extraction from recovered platforms as a primary threat vector~\cite{wang2024survey}.

Critically, an adversary who acquires credentials through these means does not necessarily also possess a flight-capable replica of the physical airframe: keys recovered from a crash site, leaked through a supply chain, or obtained via an insider carry no accompanying aircraft. Fielding a physical decoy would require evading visual, radio-frequency, or radar-based fleet verification and risking detection during the mission itself. Replaying the credential over the wireless link, by contrast, requires only proximity to the network once the legitimate node has been forced offline. The attack chain is therefore the lower-cost, lower-risk path to persistent access for an adversary who holds a key but lacks a deployable aircraft distinct from, and requiring less capability than, physically substituting a rogue drone.

\subsection{Attack Chain}

\begin{figure}[t]
    \centering
    \includegraphics[width=\columnwidth]{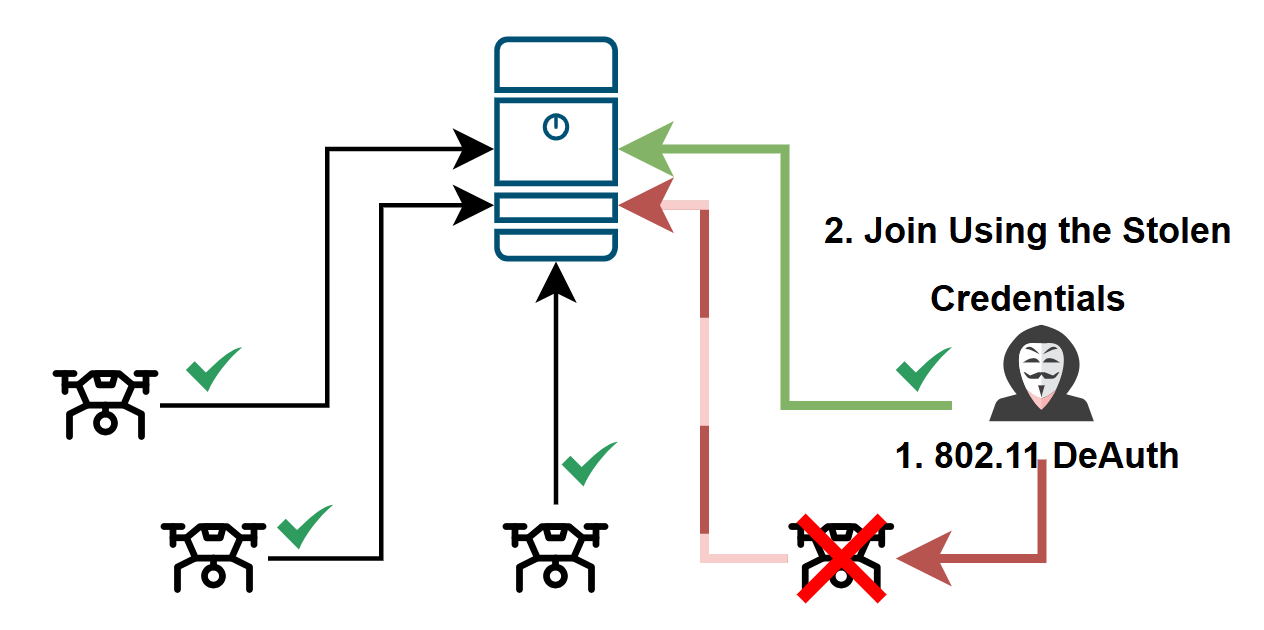}
    \caption{Chained attack workflow: Stage 1 forces disconnection via 802.11 deauthentication; Stage 2 exploits the vacancy using extracted credentials.}
    \label{fig:attack_chain}
\end{figure}

The chained attack, illustrated in Fig.~\ref{fig:attack_chain}, proceeds in two stages.

\textbf{Stage 1---Targeted Network Disruption.} Using the MAC addresses identified in Fig.~\ref{fig:targetedscan}, the adversary directs a burst of deauthentication frames at a specific drone. This action forces the drone into a persistent offline state, thereby preventing it from contributing local updates to the global model. Consequently, the aggregator does not receive the client's local updates.

\textbf{Stage 2---Credential Impersonation.} After obtaining credentials from the attacked drone or anticipating the recovery of the disconnected drone, the adversary initiates an impersonating client using the cloned credentials. Because the legitimate client is offline due to Stage 1, the FL server accepts the impersonator as valid. This occurs because single-factor authentication cannot distinguish between the original hardware and an adversary presenting identical credentials.

\subsection{Attack Objectives}

The combined attack is designed to achieve several objectives:
\begin{itemize}
    \item \textbf{Immediate degradation:} Training instability through forced disconnections.
    \item \textbf{Persistent access:} Gaining uninterrupted entry into federation rounds through impersonation.
    \item \textbf{Stealth:} Avoiding authentication failures in logs, causing the adversary to appear as a legitimate returning client.
\end{itemize}

\section{Methodology}

We validate the attack chain through two complementary experimental phases using Raspberry Pi and NVIDIA Jetson devices as drone surrogates. We adopt this compute-surrogate approach because the vulnerability under study resides in the 802.11 link layer and the FL client-authentication layer, neither of which depends on airframe dynamics, flight controller behaviour, or aerial radio propagation. Any platform running the same wireless stack and FL client software is subject to an identical attack surface at these two layers, regardless of whether it is airborne. This is consistent with prior benchmarking of edge-compute hardware for UAV payloads~\cite{minott2025benchmarking,fei_unmanned_2025}. Our approach establishes a causal link between physical-layer connectivity disruptions and the fundamental performance of EI in autonomous swarms.

\subsection{Experimental Testbeds and Hardware Configuration}

\begin{figure}[t]
    \centering
    \includegraphics[height=8cm,width=8cm]{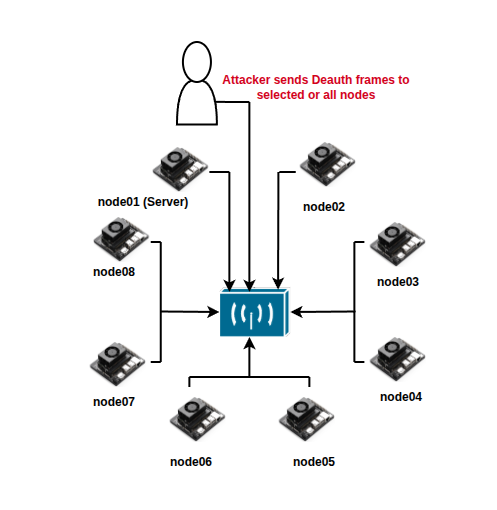}
    \caption{Experimental testbed: 8-node Jetson cluster with dedicated attacker node for deauthentication injection.}
    \label{fig:testbed}
\end{figure}

We deployed two standalone physical clusters, shown in Fig.~\ref{fig:testbed}, connected via an isolated 802.11 wireless network:
\begin{itemize}
    \item \textbf{General-Purpose Edge Cluster:} A 10-node homogeneous cluster of Raspberry Pi 4B units (Broadcom BCM2711, 4\,GB RAM), one PC workstation acting as the FL server, running the Flower framework to coordinate model broadcast and aggregation.
    \item \textbf{GPU-Accelerated Cluster:} An 8-node NVIDIA Jetson cluster. In this setup, 7 Jetson modules function as the clients, while the 8th module acts as a localized SuperLink (server) aggregator.
\end{itemize}

An additional Raspberry Pi equipped with a REALTEK monitor-mode adapter (RTL8812AU) served as the adversarial node. This node performed the reconnaissance scans visualized in Fig.~\ref{fig:targetedscan} and executed the synchronized deauthentication injections.

The isolated wireless network used in our experiments operated without 802.11w Management Frame Protection (PMF) enabled, which is why unauthenticated deauthentication frames were effective in Stage 1. We acknowledge this as a boundary condition of our threat model rather than an incidental configuration choice: PMF support remains inconsistent among several representative IoT development platforms, some of which lack full out-of-the-box support even when the underlying hardware is technically capable~\cite{alghisi2024wpa3iot}.

\subsection{FL System Configuration}

We implemented synchronous Federated Averaging (FedAvg)~\cite{mcmahan2017fedavg} using the Flower framework~\cite{beutel2020flower}. The server was configured to tolerate partial participation, enabling rounds to proceed despite significant dropouts. The CIFAR-10 dataset was used with a standard CNN architecture. The number of global rounds was set to 10 because extending training beyond 10 rounds resulted in only marginal performance improvements, while increasing runtime and adding little analytical value to the comparison. This stopping choice isolates the attack-induced convergence disruption without extending runtime.

\subsubsection{Data distributions}
\begin{itemize}
    \item \textbf{IID:} Training data shuffled and randomly partitioned equally across clients.
    \item \textbf{Non-IID:} We generated label-skewed client partitions using the Flower DirichletPartitioner over the CIFAR-10 training set, partitioning by label with concentration parameter $\alpha=0.3$ and seed $42$. Dirichlet partitioning induces both class imbalance and unequal per-client dataset sizes; in our 10-client setup, client sizes ranged from approximately 1.1k to 9.7k samples while remaining disjoint and summing to the CIFAR-10 training set size (50,000).
\end{itemize}

\subsubsection{Model Weight Verification via Network Traffic Analysis}

Model parameter transmission during federated learning was verified using a multi-stage framework that integrated packet capture, Protobuf decoding, and cross-layer log alignment. During the initial stages of our testbed deployment, all client-server communications were conducted without TLS encryption to enable direct packet inspection and validation of the FL protocol implementation. This approach enabled us to inspect the contents of network packets and verify that the Flower framework was transmitting model parameters rather than metadata, logs, or other auxiliary information.
Network traffic was captured using Wireshark, through which \texttt{PushObjectRequest} messages generated by the Flower client were identified. Each of these messages contained a 746-byte \texttt{object\_content} field encoded with Google Protocol Buffers. A custom Python decoder was implemented to interpret the payload. The decoding process identified a fully connected layer with shape $(10, 84)$. An excerpt of the recovered weight matrix is presented below:

\begin{lstlisting}[style=terminal]
Preview of weights:
[[ 1.83726791e-02 -2.09797233e-01  4.93503697e-02 -2.98543628e-02
   8.38919808e-02  6.39304482e-02 -5.98880658e-01  1.36638818e-02
  -2.08551668e-02 -3.18821784e-02 -3.88097906e-02 -1.10503078e-01
   6.86666186e-02  1.21692688e-02  0.09434399e-01 -4.29202149e-01
  ...
  [-5.10263558e-02  7.62932122e-02  1.36620075e-02  4.37011030e-02
  -6.22387230e-02  2.78727490e-02  6.46504440e-02  4.90706244e-02
  -2.68124342e-01 -5.11277068e-02 -1.31564449e-02  1.77118689e-02]]
\end{lstlisting}

\begin{figure}[t]
    \centering
    \includegraphics[width=\columnwidth]{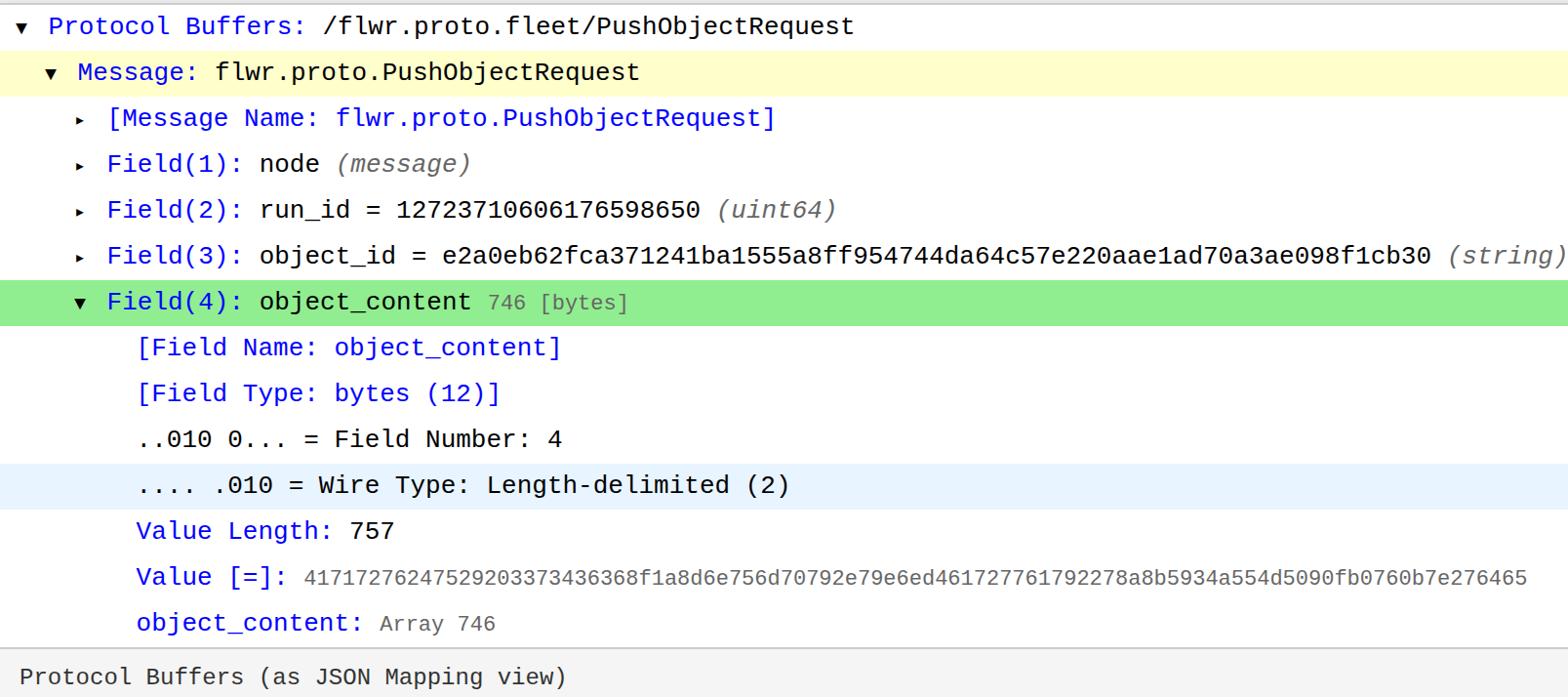}
    \caption{Wireshark capture of the Flower \texttt{PushObjectRequest} packet.}
    \label{fig:wireshark}
\end{figure}

\noindent As shown in Fig.~\ref{fig:wireshark}, the Wireshark capture demonstrates the structure of the \texttt{PushObjectRequest} packet, with the highlighted \texttt{object\_content} field clearly visible. This plaintext inspection phase served as a critical baseline for understanding legitimate FL communication patterns, essential context for our subsequent analysis of malicious model updates. After TLS was enabled for all subsequent experiments, on-wire inspection of model parameters was no longer feasible without access to session keys.

\subsubsection{Attack Injection}
Targeted deauthentication attacks were executed by the adversary using \texttt{aireplay-ng} with the following parameters:

\textbf{Intensity (I):} Percentage of clients targeted per round: $I \in \{20\%, 40\%, 60\%, 80\%\}$, corresponding to 2, 4, 6, and 8 clients respectively.

\textbf{Timing:} Attacks synchronized to FL workflow phases pre-round (blocking model download) and mid-round (disrupting training/upload).

\subsubsection{Metrics}

\begin{itemize}
    \item \textbf{Accuracy Degradation} ($\Delta A$): $\Delta A = A_{\text{baseline}} - A_{\text{attack}}$
    \item \textbf{Convergence Instability} ($\sigma_L$): Standard deviation of training loss over the attack-affected rounds
    \item \textbf{Connectivity Cost} ($C_C$): Per-round latency penalty from reconnection
\end{itemize}

\subsection{Stage 2: Credential Impersonation}

After demonstrating that deauthentication attacks can reliably disconnect legitimate clients in Stage 1, the next step is to evaluate whether an adversary can exploit this disruption window to impersonate the disconnected drone. This evaluation corresponds to the second phase of the attack chain illustrated in Fig.~\ref{fig:attack_chain}. It is assumed that the adversary has obtained valid client credentials through any available means, including physical capture, insider access, credential leakage, or supply chain compromise. The focus is on assessing the system's resilience to impersonation, rather than the method of credential acquisition.

\subsubsection{Extended Configuration}

Building on the Stage 1 infrastructure, we extended our testbed to evaluate credential-based impersonation. The FL server was configured with the following features:
\begin{itemize}
    \item TLS encryption for all communications
    \item CSV-based node authentication using EC key pairs for timestamp signing
\end{itemize}

\subsubsection{Attack Scenarios}

We evaluated two impersonation scenarios to reflect different stages of adversarial progression.

\textbf{Scenario 1 (Concurrent):} The attacker attempts to connect while the legitimate client remains active, using cloned credentials. This baseline scenario tests whether the system detects duplicate identities.

\textbf{Scenario 2 (Post-Disruption):} After a successful Stage 1 deauthentication attack, the legitimate client is disconnected. The attacker subsequently connects using the victim's credentials, thereby completing the full attack chain depicted in Fig.~\ref{fig:attack_chain}.

\section{Results}

\subsection{Stage 1: Impact of Network Disruption}

\subsubsection{IID Data Distribution (Raspberry Pi vs.\ Jetson)}

Under IID conditions, both hardware platforms demonstrated resilience to availability attacks, though with distinct absolute accuracy levels and degradation patterns. The Raspberry Pi testbed achieved a baseline accuracy of 57.00\% after 10 training rounds, which declined to 49.80\% under sustained attack (rounds 5 to 8 at escalating intensity), representing a 7.20\% reduction in final accuracy. In comparison, the Jetson testbed achieved a higher baseline accuracy of 73.40\% after 10 rounds, with the attack condition converging to 71.10\% accuracy, resulting in a 2.30\% loss. The smaller final accuracy gap observed on the Jetson platform indicates that increased computational capability and faster local convergence can partially offset the long-term effects of client unavailability, although transient degradation remains evident during the active attack period.

\subsubsection{Non-IID Data Distribution (Raspberry Pi vs.\ Jetson)}

Under Non-IID conditions, characterized by class imbalance across clients, both platforms exhibited reduced robustness relative to IID scenarios, with the Raspberry Pi testbed displaying greater vulnerability overall. On the Raspberry Pi, the baseline achieved 55.38\% accuracy after 10 rounds, while the attacked system converged to 39.63\%, resulting in a 15.75\% loss, which is more than double the degradation observed under IID conditions. For Jetson, the baseline Non-IID accuracy reached 76.27\% after 10 rounds, whereas the attacked configuration attained 68.41\%, resulting in a 7.86\% decrease. Although this relative loss is smaller than that observed on Raspberry Pi, Jetson Non-IID runs still exhibited a pronounced degradation plateau during the attack phase. These results underscore that statistical heterogeneity remains a primary factor in vulnerability to availability attacks, even on more capable hardware. Attack intensity analysis, as shown in Table~\ref{tab:combined_results_pi}, was conducted on both platforms; intensities up to 80\% are reported for the Raspberry Pi to characterise worst-case sensitivity under resource-constrained conditions. Jetson experiments evaluated attack intensities up to 60\% (rounds 5--7), as higher intensities yielded only marginal additional degradation due to faster local convergence.

The effects of the attack persisted throughout the active disruption period on both platforms. On Raspberry Pi, at the peak attack intensity of 80\% (round 8), the instantaneous accuracy gap between baseline and attacked Non-IID runs reached 18.07\% and did not fully recover by round 10. Jetson runs exhibited a comparable pattern during the attack window (rounds 5 to 7), with reduced accuracy under Non-IID attack relative to the Non-IID baseline, though the long-term gap was smaller by the end of training. These findings indicate that the breaking point under Non-IID conditions occurred at lower effective attack intensities than under IID on both testbeds, confirming that data distribution skew increases vulnerability to availability attacks.

\subsubsection{Targeted Expert Node Removal (Raspberry Pi)}

To assess the strategic implications of targeted client selection under Non-IID conditions, an additional experiment was conducted on the Raspberry Pi testbed by removing the two clients with the highest concentrations of specific classes. The Non-IID data distribution exhibited pronounced label skew: Client 8 (partition 8) contained 2,811 Class 1 samples, accounting for 29\% of its 9,671 total samples and serving as the primary source of Class 1 knowledge. Client 2 (partition 2) held 3,409 Class 6 samples (59\% of its 5,807), making it the dominant expert for Class 6. Together, these two clients represented approximately 31\% of the total training data.

Removal of these two expert nodes resulted in both a decrease in overall accuracy and impaired convergence. The baseline system with all 10 clients achieved 56.09\% accuracy after 10 rounds, whereas the expert-depleted system showed a 3.61\% reduction in final accuracy. Additionally, convergence speed declined: the baseline reached 50\% accuracy by round 4, while the system without experts required 7 rounds to achieve similar performance (49.08\%), representing a 3-round delay and a 75\% increase in time-to-competency. These findings indicate that targeted removal of expert nodes is more detrimental than random dropout at equivalent data volumes. This effect is intrinsic to federated learning rather than to the underlying wireless attack: the disproportionate impact arises because the adversary's disconnection choice interacts with the non-IID structure of the FL data partition. A generic wireless DoS attack against any 802.11 client cannot produce this class-specific degradation, since it has no analogue of ``expert'' data concentration to exploit.

\subsubsection{Convergence Instability ($\sigma_L$)}

Convergence instability, denoted as $\sigma_L$, was measured as the standard deviation of training loss during the attack-affected rounds on the Raspberry Pi testbed (rounds 5 to 8). For the baseline condition, the mean loss over these rounds was 1.3114 with $\sigma_L = 0.0168$, indicating stable convergence. Under the deauthentication attack, in which clients were forcibly disconnected and did not rejoin, $\sigma_L$ increased to 0.5393, reflecting substantially more volatile loss trajectories and disrupted optimization dynamics. These results indicate that availability attacks not only reduce final accuracy but also increase training process variability, complicating convergence guarantees in heterogeneous federated learning systems.

\subsubsection{Connectivity Cost ($C_C$)}
Connectivity cost, $C_C$, was calculated as the difference in mean accuracy between the baseline and attack conditions during the attack window (rounds 5--8) on the Raspberry Pi testbed. For the baseline, the mean accuracy over these rounds was 0.5249, while the mean accuracy under attack was 0.4361, resulting in $C_C = 0.0888$. This metric quantifies the operational performance penalty associated with client loss during network disruption. In the Jetson experiments, a similar pattern emerged: although absolute accuracies were higher, the attack rounds still exhibited a notable $C_C$, demonstrating that increased computational capacity does not fully eliminate the performance cost driven by connectivity loss.

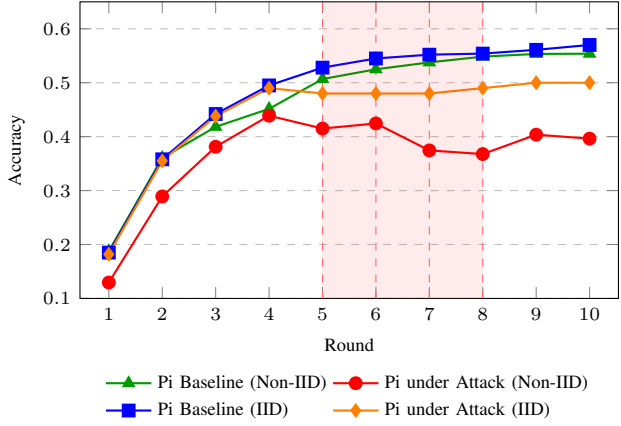
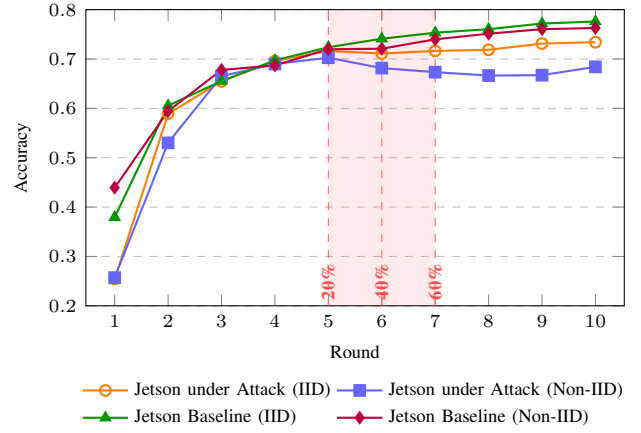
\begin{figure*}[t]
    \centering
    
    \begin{subfigure}{0.48\textwidth}
        \begin{tikzpicture}
        \begin{axis}[
            xlabel={Round},
            ylabel={Accuracy},
            xmin=1, xmax=10,
            ymin=0.1, ymax=0.65,
            xtick={1,...,10},
            ytick={0.1,0.2,0.3,0.4,0.5,0.6},
            ymajorgrids=true,
            grid style=dashed,
            width=\linewidth, 
            height=5.5cm,     
            enlarge x limits=0.06,
            tick label style={font=\scriptsize},
            label style={font=\scriptsize},
            legend style={
                at={(0.5,-0.22)},
                anchor=north,
                legend columns=2, 
                font=\scriptsize,
                draw=none
            },
            legend cell align=left
        ]
        \addplot[color=green!60!black,mark=triangle*,mark size=2.2,thick] coordinates {
            (1,0.1878) (2,0.3607) (3,0.4182) (4,0.4516) (5,0.5068)
            (6,0.5249) (7,0.5377) (8,0.5484) (9,0.5534) (10,0.5538)
        };
        \addlegendentry{Pi Baseline (Non-IID)}
        \addplot[color=red,mark=*,mark size=2.2,thick] coordinates {
            (1,0.1295) (2,0.289125) (3,0.38125) (4,0.438833) (5,0.4150)
            (6,0.4245) (7,0.3745) (8,0.367667) (9,0.403667) (10,0.396333)
        };
        \addlegendentry{Pi under Attack (Non-IID)}
        \addplot[color=blue,mark=square*,mark size=2.2,thick] coordinates {
            (1,0.185) (2,0.358) (3,0.442) (4,0.495) (5,0.528)
            (6,0.545) (7,0.552) (8,0.554) (9,0.561) (10,0.57)
        };
        \addlegendentry{Pi Baseline (IID)}
        \addplot[color=orange,mark=diamond*,mark size=2.2,thick] coordinates {
            (1,0.182) (2,0.355) (3,0.438) (4,0.49) (5,0.48)
            (6,0.48) (7,0.48) (8,0.49) (9,0.5) (10,0.5)
        };
        \addlegendentry{Pi under Attack (IID)}
        \addplot[draw=none,fill=red,fill opacity=0.08] coordinates {
            (5,0.1) (8,0.1) (8,0.65) (5,0.65)
        };
        \foreach \x in {5,6,7,8} {
            \addplot[dashed, red!60] coordinates {(\x,0.1) (\x,0.65)};
        }
        \end{axis}
        \end{tikzpicture}
        \caption{Raspberry Pi cluster accuracy.}
        \label{fig:pi_accuracy}
    \end{subfigure}\hfill
    \begin{subfigure}{0.48\textwidth}
        \begin{tikzpicture}
        \begin{axis}[
            xlabel={Round},
            ylabel={Accuracy},
            xmin=1, xmax=10,
            ymin=0.2, ymax=0.8,
            xtick={1,...,10},
            ytick={0.2,0.3,0.4,0.5,0.6,0.7,0.8},
            ymajorgrids=true,
            grid style=dashed,
            width=\linewidth, 
            height=5.5cm,     
            enlarge x limits=0.06,
            tick label style={font=\scriptsize},
            label style={font=\scriptsize},
            legend style={
                at={(0.5,-0.22)},
                anchor=north,
                legend columns=2, 
                font=\scriptsize,
                draw=none
            },
            legend cell align=left
        ]
        \addplot[color=orange,mark=o,mark size=2.0,thick] coordinates {
            (1,0.2543) (2,0.5889) (3,0.6546) (4,0.6960) (5,0.7164)
            (6,0.7110) (7,0.7163) (8,0.71856) (9,0.7312) (10,0.7340)
        };
        \addlegendentry{Jetson under Attack (IID)}
        \addplot[color=blue!60,mark=square*,mark size=2.0,thick] coordinates {
            (1,0.2570) (2,0.5301) (3,0.6652) (4,0.6899) (5,0.7025)
            (6,0.6815) (7,0.6733) (8,0.6664) (9,0.6672) (10,0.6841)
        };
        \addlegendentry{Jetson under Attack (Non-IID)}
        \addplot[color=green!60!black,mark=triangle*,mark size=2.0,thick] coordinates {
            (1,0.3794) (2,0.6055) (3,0.6551) (4,0.6973) (5,0.7238)
            (6,0.7410) (7,0.7532) (8,0.7604) (9,0.7718) (10,0.7760)
        };
        \addlegendentry{Jetson Baseline (IID)}
        \addplot[color=purple,mark=diamond*,mark size=2.0,thick] coordinates {
            (1,0.4393) (2,0.5945) (3,0.6778) (4,0.6870) (5,0.7198)
            (6,0.7208) (7,0.7398) (8,0.7512) (9,0.7604) (10,0.7627)
        };
        \addlegendentry{Jetson Baseline (Non-IID)}
        \addplot[draw=none,fill=red,fill opacity=0.08,forget plot] coordinates {
            (5,0.2) (7,0.2) (7,0.8) (5,0.8)
        } \closedcycle;
        \addplot[dashed, red!60] coordinates {(5,0.2) (5,0.8)};
        \addplot[dashed, red!60] coordinates {(6,0.2) (6,0.8)};
        \addplot[dashed, red!60] coordinates {(7,0.2) (7,0.8)};
        \node[rotate=90, text=red!70, font=\scriptsize\bfseries] at (axis cs:5,0.25) {20\%};
        \node[rotate=90, text=red!70, font=\scriptsize\bfseries] at (axis cs:6,0.25) {40\%};
        \node[rotate=90, text=red!70, font=\scriptsize\bfseries] at (axis cs:7,0.25) {60\%};
        \end{axis}
        \end{tikzpicture}
        \caption{Jetson cluster accuracy.}
        \label{fig:jetson_accuracy}
    \end{subfigure}
    
    \caption{Accuracy trajectories across training rounds under IID and Non-IID conditions for both testbeds. The red shaded regions denote active attack windows.}
    \label{fig:combined_accuracy}
\end{figure*}

Table~\ref{tab:combined_results_pi} summarises the degradation in accuracy observed on the Raspberry Pi testbed under increasing availability attack intensity for both IID and Non-IID distributions, while Table~\ref{tab:jetson_gap_summary} provides a cross-platform comparison of final-round accuracy and degradation for Raspberry Pi and Jetson systems. These results are further illustrated in Fig.~\ref{fig:combined_accuracy} (specifically Fig.~\ref{fig:pi_accuracy} for the Raspberry Pi testbed and Fig.~\ref{fig:jetson_accuracy} for the Jetson cluster), which shows the accuracy trajectories across all training rounds under both IID and Non-IID attack conditions.

\begin{table}[t]
\centering
\caption{Accuracy Degradation Under IID and Non-IID Distributions (Raspberry Pi).}
\label{tab:combined_results_pi}
\begin{tabular}{@{}cccc@{}}
\toprule
\textbf{Round} & \textbf{Attack Intensity (\%)} & \textbf{$\Delta A$ IID (\%)} & \textbf{$\Delta A$ Non-IID (\%)} \\
\midrule
5 & 20 & 4.80 & 9.18 \\
6 & 40 & 6.50 & 10.04 \\
7 & 60 & 7.20 & 16.32 \\
8 & 80 & 6.40 & 18.07 \\
\bottomrule
\end{tabular}
\end{table}

\begin{table}[t]
\centering
\caption{Cross-Platform Accuracy and Degradation (Round 10).}
\label{tab:jetson_gap_summary}
\begin{tabular}{@{}ccccc@{}}
\toprule
\textbf{Platform} & \textbf{Dist.} & \textbf{Baseline (\%)} & \textbf{Attack (\%)} & \textbf{$\Delta A$ (\%)} \\
\midrule
Raspberry Pi & IID     & 57.00  & 49.80  & 7.20  \\
Raspberry Pi & Non-IID & 55.38  & 39.63  & 15.75 \\
Jetson       & IID     & 73.40  & 71.10  & 2.30  \\
Jetson       & Non-IID & 76.27  & 68.41  & 7.86  \\
\bottomrule
\end{tabular}
\end{table}

\subsection{Stage 2: Impersonation Results}

Impersonation experiments utilized a controlled five-client federation with TLS mutual authentication and Flower node authentication enabled. Five Raspberry Pi 4 devices served as legitimate clients, and a sixth device operated as the impersonator, pre-configured with credentials extracted from Client 2 during Stage 1. The evaluation included two scenarios: concurrent impersonation while the legitimate client remained connected, and post-disconnect impersonation following forced disconnection.

\subsubsection{Attack Timeline}

The five-round training run completed in 292.04 seconds. Four distinct phases were identified from the server-side logs, with key excerpts presented below.

\paragraph{Phase 1: Normal operation (Round 1).}
All five legitimate clients were sampled, completed training, and returned results without error:

\begin{lstlisting}[style=serverlog]
[ROUND 1]
configure_fit: strategy sampled 5 clients (out of 5)
aggregate_fit: received 5 results and 0 failures
aggregate_evaluate: received 5 results and 0 failures
\end{lstlisting}

\paragraph{Phase 2: Forced disconnection (Round 2).}
The deauthentication attack began at the start of Round 2. The server initially sampled all five clients; however, midway through the round, Client 2's session was terminated, as indicated by the \texttt{DeleteNode} entry. The evaluation phase confirmed the dropout:

\begin{lstlisting}[style=serverlog]
[ROUND 2]
configure_fit: strategy sampled 5 clients (out of 5)
[Fleet.DeleteNode] Delete node_id=14308369969468497996
aggregate_fit: received 5 results and 0 failures
aggregate_evaluate: received 4 results and 1 failures
\end{lstlisting}

\noindent It is notable that the server did not log any security alert; the failure was treated as a routine client dropout.

\paragraph{Phase 3: Impersonator infiltration (Round 3).}
With Client 2 offline, the server sampled only four clients for training. During this round, the impersonator connected using Client 2's extracted private key. The server accepted the new node, assigned an identifier, and served the global model:

\begin{lstlisting}[style=serverlog]
[ROUND 3]
configure_fit: strategy sampled 4 clients (out of 4)
[Fleet.CreateNode] Created node_id=2923142953798957254
aggregate_fit: received 4 results and 0 failures
[Fleet.GetRun] Requesting 'Run' for run_id=4087803549789865333
[Fleet.GetFab] Requesting FAB for fab_hash=c65d375febbb6c398ec99...
aggregate_evaluate: received 5 results and 0 failures
\end{lstlisting}

\noindent The impersonator participated in evaluation, restoring the client count to five. No authentication failure was logged during this process.

\paragraph{Phase 4: Full integration (Rounds 4--5).}
From Round 4 onward, the impersonator was indistinguishable from legitimate participants, contributing model updates during both training and evaluation:

\begin{lstlisting}[style=serverlog]
[ROUND 4]
configure_fit: strategy sampled 5 clients (out of 5)
aggregate_fit: received 5 results and 0 failures
aggregate_evaluate: received 5 results and 0 failures

[ROUND 5]
configure_fit: strategy sampled 5 clients (out of 5)
aggregate_fit: received 5 results and 0 failures
aggregate_evaluate: received 5 results and 0 failures
\end{lstlisting}

\noindent Overall, the impersonator participated in two full rounds and contributed two model updates to the global model. Training loss converged consistently throughout (Round 1: 2.30, Round 5: 1.70), confirming that the server incorporated the impersonator's updates without detecting any anomaly.

\subsubsection{Concurrent Impersonation (Scenario 1)}

When the impersonating client attempted to connect while Client 2's session remained active, the SuperLink server rejected the connection. Flower enforces a single active session per public key, preventing duplicate participation within a training round. Although this mechanism provides basic protection against credential cloning during active sessions, it does not prevent impersonation following disconnection events, whether caused by network failure, client crash, or adversarial action, as demonstrated above.

\subsubsection{Post-Disconnect Impersonation (Scenario 2)}

The interval between disconnection and successful impersonation spanned approximately one training round (approximately 58 seconds), demonstrating that adversaries possess a practical exploitation window. From the server's perspective, the impersonator was indistinguishable from the legitimate client throughout. These results confirm that Flower's authentication model operates on a possession-based assumption: valid credentials imply legitimate identity. This approach is fundamentally flawed when credentials can be compromised through physical capture or extraction.

\subsubsection{Security Implications}

The experimental results reveal a significant gap between authentication and authorisation in federated learning frameworks. TLS mutual authentication verifies that a connecting client possesses valid credentials, but it cannot determine whether the credential holder is the originally provisioned device, whether the connection follows a legitimate or adversarial disconnection, or whether the device's integrity has been compromised. Single-session enforcement protects only against concurrent credential reuse, leaving a persistent vulnerability window after any disconnection event. In adversarial environments where physical capture is feasible, such as autonomous drone deployments, this window can be reliably created and exploited.

\section{Ethical Considerations}

All experiments were conducted on isolated, researcher-controlled Raspberry Pi and Jetson testbeds and a dedicated wireless network; no third-party systems or operational drone platforms were targeted. The study did not involve human subjects, interviews, or human-derived sensitive data. Network traces and device identifiers used during experimentation were limited to experiment-owned devices and sanitized where appropriate in the paper. In accordance with responsible disclosure best practices, the identified authentication vulnerability was reported to the affected framework's development team prior to submission.

\section{Discussion}

This study investigated a chained attack targeting drone-based FL systems, which combines network-layer denial-of-service with credential-based impersonation. The empirical results indicate that: (1) deauthentication attacks substantially degrade training performance, especially under non-IID data conditions where the loss of an ``expert'' client results in irreversible accuracy reductions; and (2) single-factor, credential-based authentication in the Flower framework allows post-disconnect impersonation to occur without detection. The attack chain, consisting of forced disconnection followed by impersonation, presents a practical threat to FL deployments in environments with physical accessibility.

\subsection{The Chained Attack Surface}

The chained attack creates a compounded threat profile that exceeds the individual impact of each component. By combining availability disruption with authentication bypass, an adversary can escalate a temporary denial-of-service into a persistent integrity breach.

\subsubsection{Forensic Invisibility}
A principal finding is the forensic indistinguishability of the attack. From the server's perspective, the victim node appears to disconnect due to routine instability, such as battery depletion or signal loss, and subsequently reconnects. Since the FL aggregator is stateless, it accepts the returning client without identifying the session gap as suspicious. This stealth characteristic ensures the attack leaves no application-layer logs, requiring correlation of physical-layer Wi-Fi data with application-layer training logs for effective anomaly detection.

\subsubsection{Targeted Model Blindness (Non-IID Condition)}
Under non-IID conditions, the availability attack becomes highly targeted. Experimental results show that disconnecting nodes leads to a disproportionate reduction in model accuracy. An adversary with knowledge of the data distribution can strategically disconnect expert nodes to achieve two objectives: (1) induce targeted class-specific failures while maintaining superficially acceptable overall accuracy metrics, and (2) significantly delay model convergence, thereby degrading operational readiness. In drone surveillance scenarios where rapid threat detection is critical, removing expert nodes for specific vehicle types or threat classes enables selective blindness attacks that evade aggregate accuracy monitoring while severely impairing performance on strategic targets.

\subsection{Generalizability and Implications}

Although empirical validation was performed using the Flower framework, the identified vulnerability constitutes a conceptual flaw in the standard FL protocol rather than an implementation-specific defect. Most FL aggregation strategies, such as FedAvg, are designed to be stateless and resilient to high client churn, implicitly assuming that valid credentials ensure valid identity regardless of session continuity.

The consequences for autonomous drone swarms are substantial. In military or search-and-rescue scenarios, the attack chain enables an adversary to degrade swarm performance and assume control of a node's identity.
\begin{itemize}
    \item \textbf{Persistence:} If credentials are static and reused across missions, a single successful key extraction allows indefinite impersonation, even after the physical device is recovered or destroyed.
    \item \textbf{Strategic Manipulation:} An adversary could suppress recognition of hostile assets by targeting and replacing drones responsible for training on those assets.
\end{itemize}

Existing defenses addressing either availability or authentication in isolation are insufficient. Comprehensive protection likely requires integrating network, authentication, and aggregation-layer defenses. Promising directions include hardware-local attestation for client-side verification, cryptographic freshness proofs for hardware continuity, and differential privacy hardening; we leave the design and evaluation of these mechanisms to future work.

\section{Conclusions}

This study has demonstrated a practical chained attack against drone-based federated learning systems that combines network-layer denial-of-service with credential-based impersonation. The empirical validation on Raspberry Pi and Jetson testbeds confirms that deauthentication attacks substantially degrade training performance, particularly under non-IID data distributions, while single-factor authentication in the Flower framework permits post-disconnect impersonation without detection.

Future work should pursue: (1) multi-trial statistical validation with larger swarms (50 to 100 clients) to quantify variance in attack effectiveness and scaling effects; (2) development and prototyping of multi-factor authentication methods suitable for resource-constrained drones; and (3) evaluation of the attack chain across a broader range of FL frameworks and drone hardware platforms.

\bibliographystyle{IEEEtran}
\bibliography{ref}

\end{document}